\documentclass[10pt,conference]{IEEEtran} 
\usepackage{cite}
\usepackage{amsmath,amssymb,amsfonts}
\usepackage{textcomp}
\usepackage{bookmark}
\usepackage{enumitem}
\usepackage{dirtytalk}
\usepackage{xcolor}

\def\BibTeX{{\rm B\kern-.05em{\sc i\kern-.025em b}\kern-.08em
    T\kern-.1667em\lower.7ex\hbox{E}\kern-.125emX}}

    \makeatletter

\begin{document}

\title{Creative Problem-Solving: A Study with Blind and Low Vision Software Professionals
}

\author{
\IEEEauthorblockN{\begin{tabular}{c} Karina Kohl \\ \textit{Institute of Informatics} \\ \textit{UFRGS}\\ Porto Alegre, Brazil \\ karina.kohl@inf.ufrgs.br \end{tabular}}
\and
\IEEEauthorblockN{\begin{tabular}{c} Yoonha Cha \\ \textit{Department of Informatics} \\ \textit{University of California, Irvine}\\ Irvine, USA \\ yoonha.cha@uci.edu \end{tabular}}
\and
\IEEEauthorblockN{\begin{tabular}{c} Victoria Jackson \\ \textit{Department of Informatics} \\ \textit{University of California, Irvine}\\ Irvine, USA  \\ vfjackso@uci.edu \end{tabular}}
\and
\IEEEauthorblockN{\begin{tabular}{c} Stacy Branham \\ \textit{Department of Informatics} \\ \textit{University of California, Irvine}\\ Irvine, USA \\ sbranham@uci.edu \end{tabular}}
\and
\IEEEauthorblockN{\begin{tabular}{c} Andr\'e van der Hoek \\ \textit{Department of Informatics} \\ \textit{University of California, Irvine}\\ Irvine, USA \\ andre@ics.uci.edu \end{tabular}}
\and
\IEEEauthorblockN{\begin{tabular}{c} Rafael Prikladnicki \\ \textit{School of Technology} \\ \textit{PUCRS}\\ Porto Alegre, Brazil \\ rafael.prikladnicki@pucrs.br \end{tabular}}
}

\maketitle
\begin{abstract}
\textit{Background}: Software engineering requires both technical skills and creative problem-solving. Blind and low-vision software professionals (BLVSPs) encounter numerous workplace challenges, including inaccessible tools and collaboration hurdles with sighted colleagues.  
\textit{Objective}: This study explores the innovative strategies employed by BLVSPs to overcome these accessibility barriers, focusing on their custom solutions and the importance of supportive communities.  
\textit{Methodology}: We conducted semi-structured interviews with 30 BLVSPs and used reflexive thematic analysis to identify key themes.  
\textit{Results}: Findings reveal that BLVSPs are motivated to develop creative and adaptive solutions, highlighting the vital role of collaborative communities in fostering shared problem-solving.   
\textit{Conclusion}: For BLVSPs, creative problem-solving is essential for navigating inaccessible work environments, in contrast to sighted peers, who pursue optimization. This study enhances understanding of how BLVSPs navigate accessibility challenges through innovation.  
\end{abstract}


\section{Introduction}



In software development, creativity and problem-solving have been found to be crucial to overcome challenges resulting from its inherent complexity \cite{glass_software_2006, brooks_man_month}, in addition to technical expertise \cite{Groeneveld2021}. 
The value of creative problem-solving manifests in solving unsolved problems, improving existing solutions, or enhancing efficiency~\cite{Wang2022}.

Within organizations, creativity results from individual and contextual factors ~\cite{George2001, Denning2012}. Software researchers have noted that personality traits such as openness to experience can predict a developer's creativity intention \cite{amin_impact_2020}. Equally important is fostering an environment that supports creativity, such as minimizing interruptions to enable the flow state crucial for individual innovation \cite{ritonummi_flow_2023}, or benefiting from a leader who employs an ambidextrous leadership style \cite{dasilvaHowSoftwareDevelopment2016}.

In recent years, the software profession has recognized its diversity crisis \cite{albusays_diversity_2021}, and research about disability in software engineering remains relatively scarce compared to other factors of diversity, equity, and inclusion \cite{cha_career_2024}. Among the sparse literature are studies about Blind and Low Vision Software Professionals (BLVSPs) which document the challenges that arise due to inaccessible tools and ableist attitudes of colleagues \cite{pandey_understanding_2021, cha_career_2024}. As an example, many programming tools have accessibility flaws, such as hard-to-use Integrated Development Environments (IDEs) that hinder code navigation~\cite{Mealin2012}. Additionally, screen readers ---assistive technologies that convert visual displays into audio or braille --- can place a significant cognitive burden on BLVSPs ~\cite{Armaly2018b}. 

Heeding the call for more research on disability in Software Engineering~\cite{cha_career_2024}, this study follows up on prior work in which we conducted semi-structured interviews with 30 BLVSPs about Do-It-Yourself (DIY) solutions they built and used to deal with workplace inaccessibility \cite{cha_chi_2024}. While analyzing the data, we identified creative problem-solving as an underlying theme in the responses. To the best of our knowledge, this aspect remains understudied in addressing workplace accessibility issues. This study presents results from a re-analysis of the interviews, explicitly focusing on creativity as the central theme while drawing upon established frameworks for understanding creative approaches. We asked the following research question:



\textit{\textbf{How do blind and low vision software professionals employ creative problem-solving strategies to adapt development tools and environments to enhance productivity?}}


This study contributes to the field in the following ways:
\begin{itemize}
\item Investigating and showcasing the creative problem-solving strategies employed by BLVSPs.
\item Offering insights into how BLVSPs adapt to and overcome accessibility challenges.
\item Bridging the domains of software development creativity and accessibility research by highlighting the distinctive workflow patterns of BLVSPs.
\end{itemize}
\pagebreak


\section{Background and Related Work} \label{background}

In this section, we provide background and related work on creativity and Assistive Technologies (AT) in software engineering, as well as their relevance to Blind and Low-Vision Software Professionals (BLVSPs).

\textbf{Creativity.} Researchers commonly define creativity through novelty and usefulness~\cite{runco_standard_2012}. In this paper, we adopt Amabile's definition of creativity as \say{The production of novel and useful ideas by individuals or small groups working together}~\cite{Amabile2016}. While there are other models to explain creativity in an individual, e.g.,  the systems view of creativity~\cite{csikszentmihalyi_creativity_2013}, we use Amabile's componential model~\cite{Amabile2016} that emphasizes the importance of an individual's domain skills, intrinsic and synergistic extrinsic motivation, and the social environment surrounding them in fostering creativity. 


\textbf{Assistive Technologies (AT)} support people with disabilities in maintaining or improving various functions~\cite{WHO_2023}. BLV individuals primarily use screen readers (e.g., JAWS~\cite{jaws}
, NVDA~\cite{NVDA}
) to provide speech synthesis for screen navigation and screen magnifiers to enable zooming for those with usable vision. These tools are essential for BLVSPs in coding and communicating~\cite{Mealin2012, Amin2024}. 

\textbf{Creativity in Software Engineering.} Research shows developers view creativity as generating innovative, effective solutions through technical skills and problem-solving~\cite{Groeneveld2021} that emphasizes clever reuse and usefulness over novelty \cite{Inman2024}. Studies have explored various aspects, from workshop formats~\cite{maiden_creative_2006} to design thinking~\cite{parizi_how_2022}. Aligned with Amabile's model, research shows personality traits affect creative intention~\cite{amin_impact_2020}, while happiness improves problem-solving~\cite{Graziotin2014} and creativity~\cite{graziotin_what_2018}. Environmental factors like supportive management~\cite{monteiro_innovative_2016} and flow states~\cite{ritonummi_flow_2023} enhance creativity.

\textbf{Blind and Low Vision Software Professionals.} BLVSPs face significant challenges in both individual and collaborative tasks~\cite{meyer_developers_2017, stray_coordination_2020} due to inaccessible tools and ableist attitudes, requiring additional access labor~\cite{branham_invisible_2015, cha_participate_2024}. These challenges are exacerbated by inaccessible development tools and limited accessibility consideration in collaborative practices~\cite{Mealin2012}.

Research has identified various barriers, particularly in programming-related tasks~\cite{mountapmbeme_addressing_2022}. IDEs like Visual Studio Code~\cite{vscode} 
and command line interfaces (CLIs) pose challenges in code comprehension and navigation~\cite{armaly_comparison_2018}, while pair programming presents difficulties due to limited AT availability on colleagues' machines~\cite{pandey_understanding_2021}. Beyond coding, BLVSPs encounter challenges in design tasks, documentation, and meetings~\cite{pandey_understanding_2021, cha_participate_2024}. Teamwork experiences significantly impact workplace accessibility~\cite{Mealin2012}, though many hesitate to request sighted assistance~\cite{storer_IDE_2021}. These challenges affect career mobility, with some avoiding job changes due to established accessibility accommodations~\cite{cha_career_2024}.

To the best of our knowledge, our study is the first to explore creativity in the workplace experiences of BLVSPs. As such, it complements previous research on software developers' workplace experiences with creativity, such as the studies by Groeneveld et al.\cite{Groeneveld2021} and Inman et al.\cite{Inman2024}.

\section{Methodology} \label{methodology}
\subsection{Study Design}
The interview protocol \cite{Kohl2025_Protocol} was developed based on prior studies of software engineering and the experiences of individuals with varying levels of vision. Questions were designed to elicit in-depth insights into participants' experiences and viewpoints, focusing on challenges and strategies employed by individuals with different visual abilities.

\subsection{Data Collection}
We conducted semi-structured interviews~\cite{Seaman2008} with 30 BLVSPs who worked in software engineering positions in corporate settings or as freelancers. Participants were recruited through professional contacts, specialized mailing lists like Program-L, and snowball sampling. 

Interviews were conducted via Zoom between May and July 2024, with audio recorded. Participants reviewed study materials and completed a demographic survey before interviews. Questions focused on workplace accessibility challenges, use of existing solutions, tool development, and experiences sharing tools with colleagues. Each participant received \$40/hour compensation via gift card. The study was approved by the researchers' Institutional Review Board (IRB).

Demographics included 18 (60\%) totally blind participants and 12 (40\%) with varying visual abilities. Participants were from the USA (18), Europe (6), India (4), and Brazil (2), with 26 men, 3 women, and 1 non-binary participant.

\subsection{Data Analysis}
All interviews were transcribed and quality-checked by authors, with Portuguese interviews translated to English. We used abductive analysis \cite{Thompson2022}; we inductively identified some creativity-related themes from the data and were also informed by Amabile's model of creativity to deduce some themes.


Researchers conducted a two-phased data analysis. First, three researchers independently coded seven randomly selected transcripts, developing a formal codebook through consensus. Second, two researchers coded the remaining 23 transcripts, frequently meeting to reconcile findings and discuss emerging themes. Previously coded transcripts were revisited when new patterns emerged to maintain consistency.

\subsection{Researcher Reflexivity}
Our multi-disciplinary research team had expertise in both Software Engineering and Human-Computer Interaction (HCI) expertise, with researchers experienced in studying people with disabilities in professional contexts and industry software development. This diversity enabled crafting culturally sensitive questions, building rapport through technical knowledge, and analyzing data through complementary lenses of software development and accessibility.

To mitigate potential biases, we implemented a collaborative analysis involving three researchers, ensuring findings emerged from multiple viewpoints rather than single interpretations. This triangulation strengthened validity while maintaining sensitivity to participants' experiences.

\renewcommand{\arraystretch}{1.5}
\begin{table*}[t!]
\centering
\caption{Themes and Main Findings}
\label{tab:themes}
\begin{tabular}{|l|l|}
\hline
\textbf{Themes}      & \textbf{Description}             \\ \hline
\textbf{Motivation Driven by Need}      & Necessity sparks creative problem-solving and drives autonomy             \\ \hline
\textbf{Creation of Opportunities}      & Personal challenges transform into broader organizational innovations     \\ \hline
\textbf{Creation of Tools}              & Highlights how BLVSPs develop custom solutions to enhance workplace autonomy  \\ \hline
\textbf{Use and Customization of Tools} & Demonstrates their technical expertise in adapting existing tools              \\ \hline
\textbf{Workflow Strategies}            & Shows how they creatively modify processes rather than creating new artifacts \\ \hline
\textbf{Communities, Sharing, Publicize}        & Emphasizes their commitment to open-source collaboration and solution sharing  \\ \hline
\end{tabular}
\end{table*}
\renewcommand{\arraystretch}{1}

\section{Findings} \label{results}

The following subsections present the themes and findings addressing our research question, summarized in Table \ref{tab:themes}.

\subsection{Motivation Driven by Need}
Research on blind and low-vision professionals reveals how necessity often catalyzes creative problem-solving, mixing with intrinsic motivation as conceptualized by Amabile \cite{Amabile2016}. The drive for independence and autonomy emerged as a motivator as illustrated by one participant who emphasized, \say{\textit{Personal autonomy, independence, and being able to do my work well, and more efficiently. I think primarily, it's really independence and autonomy}} (P12). 

For our participants, the development of custom solutions originated from personal experience with inadequate existing tools. As one participant noted, \say{\textit{A few of these add-ons were developed because we were complaining about something in our professional lives that didn't work well}} (P7). This observation resonates with Amabile's \cite{Amabile2016} research on the progress principle, where the intrinsic motivation to create solutions is heightened when individuals encounter obstacles that directly impact their daily work progress. This is evident in cases where participants had to adapt to vision loss without prior experience or guidance, as exemplified by one participant who shared, \say{\textit{I've never been blind before. I don't know anybody who's blind... So I was left on my own to find stuff}} (P22). The frustration with unreliable solutions drove them to develop more dependable tools, demonstrating how necessity fosters intrinsically motivated creative problem-solving.

The complexity of working with existing ATs, as screen reader scripts, illustrates how technical challenges can spark creative engagement. One participant highlighted: \say{\textit{These [screen reader] scripts are very, very complex. They make it look easy, but it's kind of complex to implement}} (P20).

\subsection{Creation of Opportunities}
 
Our data reveals how BLVSPs turned personal challenges into opportunities for collective innovation. One participant mentioned, \say{\textit{You just have to frame the problem in a way that benefits the rest of the team}} (P24). This highlights how their 
motivation to address accessibility challenges goes beyond personal needs, encompassing broader organizational benefits.

For example, P12, who is part of the accessibility and innovation team, stated that their role involved \say{\textit{working closely with [the company's] features and trying to leverage the latest technology for accessibility scenarios.}} This demonstrates how intrinsically motivated professionals can effectively connect cutting-edge technology with accessibility needs.

\subsection{Creation of Tools}

The development of custom tools by BLVSPs exemplifies how motivation drives creative exploration and problem-solving in professional contexts. Drawing on Amabile's \cite{Amabile2016} theory of creativity, we observed how BLVSPs' motivation to improve their work environment leads to innovative solutions that address fundamental accessibility challenges. This is particularly evident in their approach to screen reader optimization, where one participant identified a core issue: \say{\textit{The current screen readers just read all this gibberish that doesn't even help. It just sprays everything it can find... the screen readers don't really understand the structure of the document that they're reading}} (P22). 

The creation of specialized tools extends beyond basic accessibility needs to address nuanced professional challenges, particularly in maintaining workplace autonomy and professional presence. As illustrated by one participant's development of a video positioning tool: \say{\textit{I was going into a video meeting and I had to ask whether I was in the shot of my webcam or not. I was worried about what me asking was doing to people's perceptions of my abilities.}}

\subsection{Use and Customization of Tools}

BLVSPs use their technical expertise to adapt tools and overcome accessibility barriers, maintaining competitive productivity levels. As one participant noted, \say{\textit{I care a lot about the pace that I can do things at. It’s not about my technical knowledge, but I’m slower at many tasks, which can affect how others perceive pairing with me}} (P24). This awareness drives them to create and implement novel solutions that enhance efficiency while upholding high standards.

Adaptations often focus on customizing tools, such as the IndentNav NVDA plugin, which improves code navigation by enabling movement between code blocks: \say{\textit{I sing its praises all the time. IndentNav lets me move between blocks at the same indentation level}} (P7). Similarly, the Phonetic Punctuation NVDA add-on replaces characters with sounds to save time: \say{\textit{When you’re reading something thousands of times, small changes can save a lot of time}} (P24). These examples align with Amabile's \cite{Amabile2016} emphasis on domain-relevant skills in creative problem-solving.

Complex challenges often demand advanced technical solutions, as illustrated by a participant who made Visual Studio's autocomplete functionality accessible: \say{\textit{Autocomplete didn’t work with screen readers, so I had to write everything by hand}} (P10). His solution, which bridged gaps in multiple environments, showcases how domain expertise enables BLVSPs to innovate and create accessible workflows, reflecting both Amabile's \cite{Amabile2016} framework and Inman et al.'s \cite{Inman2024} concept of creativity through clever reuse where existing functionalities are reimagined and repurposed to create more accessible and efficient workflows.

\subsection{Workflow Strategies}

For BLVSPs, creativity manifests primarily through adaptive problem-solving and strategic workflow development rather than traditional artifact creation. Considering Amabile's \cite{Amabile2016} theory of creativity, these professionals demonstrate strong domain skills in accessibility technologies, coupled with process-relevant skills in identifying and implementing workarounds, driven by the motivation to overcome accessibility barriers. This creative adaptation also aligns with the clever reuse by Inman et al. \cite{Inman2024} where existing tools and processes are repurposed to serve new accessibility needs, well-illustrated by P24's comment where he described that he lives \say{\textit{in a [sighted] world that hasn't been designed for people like [me],}} thus having to come up with creative ways to use inaccessible products and tools from a very young age.. 


The nature of these creative solutions often manifests in strategic rather than physical innovations. For instance, P5 described developing specific \say{\textit{tricks}} for window management and email organization: \say{\textit{I set some techniques... that over time I managed a way to be more effective to switch between windows if there are multiple windows opened.}} BLVSPs believed that \say{\textit{what allows me to compete with my sighted colleagues is a real ability to problem solve. To really figure out what the problem is and how I’m going to solve it, whether it’s building tools, asking for help where I need it, whatever it is}} (P25). This approach demonstrates how BLVSPs leveraged their domain expertise to create efficient workflows rather than build new tools from scratch. Their creative process often involves a careful balance between technological and human-assisted solutions, as illustrated by one participant who noted a preference for human-assisted support when dealing with inaccessible user interfaces, particularly when elements \say{\textit{aren't focusable, or aren't even in the accessibility tree}} (P28).

\subsection{Communities, Sharing, Publicize}

The motivation to share solutions arises from a shared understanding of accessibility challenges, as one participant expressed: \say{\textit{I want to get blind people together so they don’t feel alone and so we can create apps that are first-class for blind people}} (P22). This prosocial motivation, paired with the practical need for accessibility solutions, fosters creativity through reuse and adaptation \cite{Inman2024}.

Community-driven efforts, like the open-source NVDA screen reader—initiated by \say{\textit{two blind guys who couldn’t afford \$1,000 for a screen reader}} (P7)—illustrate how necessity and collaboration can lead to impactful innovation through domain expertise and creative processes \cite{Amabile2016}. Tools are further shared and refined via platforms like Program-L, with one participant encouraging broad dissemination: \say{\textit{Spread it around to as many people as possible}} (P13).

However, community collaboration also reveals challenges. While some developers see organic growth, such as gaining \say{\textit{30-some stars on GitHub}} (P7), others hesitate to share their work, fearing it is incomplete or not polished enough: \say{\textit{They didn’t look nice enough to share}} (P10). This highlights the need for structured resources, as one participant proposed: \say{\textit{There are needs to be a site like Apple Vis more for blind developers... where you can refer people to a website and all of this stuff is magically shown, as far as resources}} (P17). Overcoming such barriers is vital to supporting BLVSPs in their creative problem-solving journey.


\section{Discussion} \label{discussion}

A distinctive aspect of LVSPs is that creative problem-solving is not an optional enhancement to their work process, as it might be for sighted professionals, but rather a fundamental necessity for basic job function. While sighted professionals might employ creative solutions to optimize or improve their workflows, the drive for independence and autonomy emerges as a motivator for BLVSPs who engage in creative problem-solving simply to access and utilize standard tools and complete routine tasks.

Mandatory creativity poses a unique challenge for BLVSPs, requiring constant engagement with Amabile's \cite{Amabile2016} creativity components: domain-relevant skills, creativity-relevant processes, and task motivation to overcome accessibility barriers. This necessity-driven innovation fosters a self-reinforcing cycle, as solutions that enhance independence and efficiency further motivate creative problem-solving.


The resulting solutions often involve clever reuse of existing tools and the creation of novel workflows rather than physical innovations, aligning with Inman et al.'s \cite{Inman2024} observations about the nature of clever reuse in professional contexts. This critical analysis of existing tools demonstrates what Inman et al. \cite{Inman2024} describe as creativity through exploration, where professionals engage in deep problem analysis to identify opportunities for meaningful innovation. The creative adaptation and customization of tools by BLVSPs demonstrate the intersection of domain-relevant skills \cite{Amabile2016} and clever reuse strategies \cite{Inman2024} in professional software development.

Creating specialized tools extends beyond basic accessibility needs to address professional challenges, particularly in maintaining workplace autonomy and professional presence. This example aligns with both Amabile's \cite{Amabile2016} emphasis on motivation as a driver of creative problem-solving and Inman et al.'s \cite{Inman2024} framework of creativity through exploration, where professionals actively seek out novel solutions to enhance their work experience. The resulting tools developed by BLVSPs, such as face recognition systems for autonomous positioning guidance, demonstrate how creative exploration driven by motivation can lead to solutions that address immediate practical needs and enhance professional independence and dignity.

BLVSPs also demonstrate how creative problem reframing can catalyze organizational change. This creative reframing of accessibility challenges as opportunities for innovation reflects what Amabile \cite{Amabile2016} describes as progress-oriented motivation, where individuals are driven not just by personal necessity but by the potential to create meaningful impact across their organizations. BLVSPs transform potential barriers into opportunities for collective advancement, fostering an environment where inclusive design becomes integral to organizational success rather than a separate consideration.

The intersection of community engagement and creative problem-solving among BLVSPs represents what Amabile and Pratt \cite{Amabile2016} describe as prosocial motivation enhancing creativity. Unlike their sighted counterparts, for whom creative problem-solving might be an optional path to workflow optimization, BLVSPs face it as a daily necessity for basic professional functions. This fundamental necessity drives communities to serve as crucial catalysts for creative solutions, functioning as support networks and platforms for collaborative innovation. The social environment's role in creativity brings collective engagement and enhances both individual and group creative outcomes \cite{Amabile2016}.

The non-optional nature of creative problem-solving for BLVSPs makes open-source platforms crucial, as they provide access to tools and solutions that might otherwise be inaccessible or prohibitively expensive. Open-source facilitates what Amabile and Pratt \cite{Amabile2016} describe as meaningful work engagement, where creators see their solutions benefiting others while receiving community feedback and improvements.

What is unclear, however, is the exact nature of the motivation for the creative problem-solving. Amabile's model recognizes that some extrinsic motivators (so-called synergistic extrinsic motivators) can benefit creativity alongside intrinsic motivation. It is unclear to what extent the motivation of our participants is driven by pure intrinsic motivation (e.g., the drive to craft a workaround) or extrinsic motivators (e.g., career recognition). It may well vary depending on the individual and the environment in which they are working.


We discuss implications for practice, recommending that software development organizations allocate dedicated time and resources for BLVSPs to adapt tools and customize workflows, as creative problem-solving is key to their work. Organizations should support open-source development and enable workplace autonomy for implementing accessibility solutions. Tool developers and platform providers are encouraged to prioritize customizability, support user-created solutions, and document successful accessibility adaptations with practical examples. Software communities should strengthen open-source accessibility initiatives, establish channels for sharing solutions and feedback, and foster collaboration between BLVSPs and sighted professionals to improve tool accessibility.

\textbf{The study helps empower BLVSPs by highlighting how creative problem-solving and tool adaptation contribute to their success. The insights can help organizations integrate accessibility solutions into standard practices while fostering community collaboration between BLVSPs and sighted professionals. Tool developers can leverage these findings to create more adaptable solutions, while organizations can use this framework to align accessibility initiatives with broader institutional goals.}

\section{Threats to validity} \label{validity}




\textit{Credibility.} Our study is part of broader research on BLVSP workplace challenges. We built trust through community engagement and structured interviews, with researchers collaboratively developing codes and themes.

\textit{Transferability.} We aimed for transferability by describing the BLVSP context and their creative problem-solving processes, allowing readers to assess the applicability of our findings to other accessibility settings. We documented diverse workplace environments and technological setups to support meaningful cross-context comparisons\cite{guba_criteria_1981}.

\textit{Data Saturation.} Our 30 participants exceed the typical median (13) in BLV accessibility research~\cite{Mack_accessibility_2021}. While full saturation was not reached given the small population, participants' diverse backgrounds provided robust insight into creative problem-solving themes.

\section{Conclusion and Future Work} \label{conclusion}
 

We used a qualitative approach to explore creative problem-solving among BLVSPs. Through 30 semi-structured interviews, we identified six themes on how creative problem-solving appears in their workflows, especially in adapting and customizing solutions when existing tools fall short: (1) Motivation is Driven by Need, (2) Creation of Opportunities, (3) Creation of Tools, (4) Use and Customization of Tools, (5) Workflow Strategies, (6) Communities, Sharing, Publicize. Participants highlighted their practices of repurposing solutions and a strong commitment to sharing resources within their communities.

This research contributes to a broader effort to deepen our understanding of the experiences of BLVSPs and their innovative approaches to workplace challenges. For future work, we see value in a longitudinal study focused on the creativity of BLVSPs, examining both individual problem-solving strategies and the role of organizational culture in fostering accessible work environments.

\section{Acknowledgements} \label{acknowledgements}

We thank our participants for their valuable insights and contributions to this study. van der Hoek was supported by NSF grants CCF-2210812 and 2326489; Branham by NSF awards 2326023 and 2211790; Prikladnicki by CNPq (Brazil) and Google's 2023 AIR program. This study was financed in part by the Coordenação de Aperfeiçoamento de Pessoal de Nível Superior - Brasil (CAPES) - Finance Code 001.

\bibliographystyle{IEEEtran}
\bibliography{biblio}

\vspace{12pt}

\end{document}